\documentclass[twocolumn,showpacs,pre]{revtex4}

\usepackage{graphicx}

\begin{document}
\title{Evidence for ``fragile'' glass-forming behavior in the
relaxation of Coulomb frustrated three-dimensional systems.}

\author{M. Grousson, G. Tarjus and P. Viot}\affiliation{
Laboratoire de Physique  Th{\'e}orique des Liquides,  Universit{\'e} Pierre et
Marie Curie 4, place Jussieu 75252 Paris Cedex 05 France}
\email{viot@lptl.jussieu.fr}

\begin{abstract}
We  show by  means    of    a  Monte Carlo simulation    study    that
three-dimensional  models  with  long-range  frustration  display  the
generic phenomena seen in  fragile glassforming liquids. Due to  their
properties (absence of quenched disorder, physical motivation in terms
of structural  frustration,   and tunable  fragility),  these  systems
appear as promising   minimal  theoretical models for  describing  the
glass transition of supercooled liquids.
\end{abstract}
\maketitle
                 
The viscous  slowing down of  supercooled liquids  that leads to glass
formation  when  the temperature  is sufficiently  lowered,  remains a
challenging problem of liquid-state and condensed-matter physics.  The
phenomenon seems dramatic enough (a change by $15$ orders of magnitude
of the viscosity $\eta$ and the primary ($\alpha$) relaxation  time $\tau_\alpha $ for
a mere decrease of the temperature $T$ by a factor of $2$) and general
enough  to call for a  universal explanation.  Yet,  the search for an
underlying  universality is thwarted  by    the absence  of   observed
criticality, e.g., by the absence of  a diverging, or rapidly growing,
static susceptibility or correlation length\cite{EAN96}.

If one accepts  the premise   that  the distinctive  features of  most
glassforming liquids, namely the rapid super-Arrhenius increase of $\eta$
and $\tau_\alpha$ with decreasing $T$ and the non exponential character of the
relaxation functions, result  from collective or cooperative behavior,
progress could be made by identifying the key physical ingredients and
formulating a {\it    minimal   theoretical model } based    on  these
ingredients, in much the same way as spin  glasses have been described
in  terms of  quenched disorder and  frustration  and modelled  by the
Edwards-Anderson Hamiltonian.  One    such attempt has been   recently
made\cite{KKZNT95},     based    on    the   concept   of   structural
frustration\cite{NelsonS89}.   Such    frustration  arises because   a
locally preferred  structure,   i.e., an arrangement    of neighboring
molecules that minimizes some local free energy, cannot tile the whole
space,    because  of   the  global   constraint\cite{NelsonS89}.  The
canonical example is  a $3$-dimensional system of  spherical particles
whose locally preferred structure,   an  icosahedral cluster of   $13$
particles, cannot  form  a  pure   crystal  because of  the   $5$-fold
rotational  symmetry of   the   icosahedron\cite{NelsonS89}.   In  the
frustration-limited    domain    theory   (FLDT)    of     supercooled
liquids\cite{KKZNT95},  structural frustration  is implemented  as the
competition between  effective interactions  acting on  very different
length  scales: a short-range ordering  term that favors the extension
of the locally preferred structure and  a weak, but long-range ($1/r$)
frustrating   term that generates      a strain free  energy   growing
super-extensively  with the linear size  of the ordered region.  These
ingredients are most  simply incorporated in three-dimensional Coulomb
($1/r$) frustrated lattice models with the following Hamiltonian,
\begin{equation}\label{eq:hamilton}
H=-J \sum_{<ij>} {\bf S}_i{\bf.}{\bf S}_j+\frac{Q}{2}\sum_{i\neq j}\frac{{\bf S}_i
{\bf .}{\bf S}_j}{|{\bf r}_{ij}|},
\end{equation}
where  $J$ and $Q$ are both  positive and denote  the  strength of the
ordering  and frustrating interactions, respectively,  $<ij>$ is a sum
over  distinct pairs of nearest  neighbors and $|{\bf r}_{ij}|$ is the
distance  between sites $i$ and $j$  (the lattice spacing  is taken as
unit length);  the tunable ratio   $Q/J$ characterizes the frustration
strength.  In such models, glassiness has  to be self-induced and does
not   result from quenched    randomness   or from  specific   kinetic
constraints.

In this  letter, we present an extensive  Monte Carlo simulation study
of Coulomb frustrated  models  on a cubic lattice.   We  focus  on the
relaxation to   equilibrium and  we  do  not   consider at this  stage
out-of-equilibrium dynamics and aging phenomena in a non-equilibrated,
glassy  state.  We  show that:  (i) the  relaxation displays the  main
features  characteristic   of  the     slowing  down  of   ``fragile''
glassforming   liquids  (non  exponential decay     of the  relaxation
function, super-Arrhenius activated $T-$dependence of $\tau_\alpha$); (ii) the
``fragility'' of  the   system, e.g.,  its  degree of  departure  from
Arrhenius behavior,  increases  as the frustration  decreases, thereby
providing a description that spans the whole range from strong to very
fragile  glassformer  via  tuning of   the relative  amplitude  of the
frustrating  interaction; (iii)  the slowing  down  occurs without the
rapid growth of a correlation length and  does not follow the standard
critical   slowing down  pattern;  (iv)   the  results, including  the
existence of a  crossover temperature in  the vicinity of the critical
point of the unfrustrated system, confirm the scenario proposed by the
FLDT.

In using Eq.~(\ref{eq:hamilton}), we  have chosen spin variables ${\bf
S}_i$ that can  take several  orientations;  most of the results  have
been obtained for    the $5$-state clock model  (${\bf   S}_i=(\cos(2\pi
\theta_i/q),\sin(2\pi \theta_i/q))$, with $\theta_i$  the  orientation  of the  planar
spin and $q=5$) and the Ising model ($q=2$), but  we have also studied
the  $q=11$ state clock model. 

In  the absence of  frustration  ($Q=0$), the  models  have a critical
point   at   a  temperature   $T_c^0$,      below  which  they     are
ferromagnetic. The long-range,  coulombic interaction requires that the
total magnetization of the system be zero in order  to ensure a proper
thermodynamic  limit.   Therefore, long-range  ferromagnetic  order is
prohibited at   all  $T's$ for any   nonzero   value  of  the
frustration parameter $Q/J$;  in the FLDT,  this mimics the effect  of
structural frustration that forbids any  long-range order based on the
periodic repetition  of the  locally preferred  structure.  The models
with discrete orientations studied  here can still form ordered phases
characterized      by      modulated  patterns     (e.g.,     lamellar
phases)\cite{Dzu93},  but as shown  recently by Monte Carlo simulation
of the Ising model\cite{GTV01b},  the transition  between paramagnetic
and modulated  phases is first-order.  Following Brazovskii\cite{B75},
this result can  be interpreted  on  the basis  of the self-consistent
Hartree    approximation    which   predicts   the   occurence   of  a
fluctuation-induced  first-order   transition,  a transition   with no
nearby low-$T$ spinodal. All  the simulations  discussed below
are made in the disordered (paramagnetic) phase.  Since the transition
to the modulated  phases at $T_{DO}(Q)$  is first-order, one could  in
principle supercool  the  paramagnetic phase  to  lower  temperatures;
however, we have  found that the  lattice sizes achievable in practice
are too small to allow for a proper supercooling below $T_{DO}$.

\begin{figure}[t]

\resizebox{8.5cm}{!}{\includegraphics[angle=270]{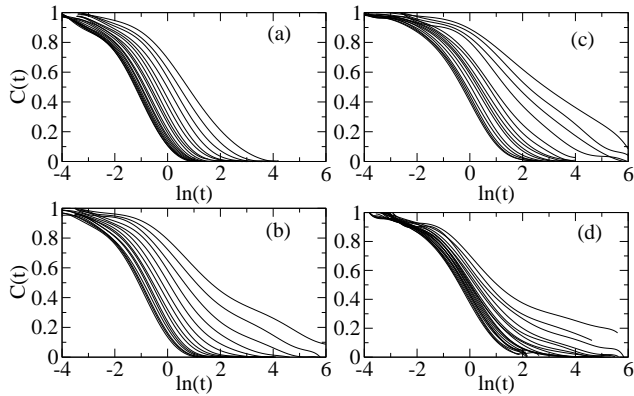}}
\caption{Spin-spin correlation function $C(t)$ versus $\ln(t)$ for
several temperatures:  $5$-state clock model for $Q=0.1$ (a) and
$Q=0.00625$ (b),  $11-$state clock model for $Q=0.05$ (c) and Ising
model for $Q=0.001$ (d). In all cases, curves from left to right are
for decreasing T's.}\label{fig:1}
\end{figure}

To study the relaxation of  the frustrated systems, we have considered
the dynamics associated with  the   Monte Carlo algorithm (simple   or
slightly  modified  Metropolis  rule\cite{Grous2001}), time  being the
number of sweeps  per spin.  In  all cases, the total magnetization is
constrained to     be zero, as   it  should  be in   the thermodynamic
limit. The dynamical quantity that we have  monitored is the spin-spin
self      correlation     function, $C(t)=(1/N)\sum_i<{\bf   S}_i(t'){\bf
S}_i(t'+t)>$, where the bracket denotes the thermal average and $N$ is
the total  number of lattice  sites.    We stress  that  $C(t)$ is  an
equilibrium correlation function that is  computed once the system has
reached   equilibrium.   In  practice,  the  thermal  average has been
performed as  an average over  $20$ different initial times  $t'$, all
chosen longer  than the relaxation  time.   Most simulations have been
done on cubic lattices of size $16^3$ to $22^3$ with periodic boundary
conditions,  and  the Coulomb  interaction has been  handled via Ewald
sums.   In  one   special  case  (see  below),  we   have  performed a
finite-size study with lattices of linear size $L=5, 10, 15$ and $20$.
We have covered two  orders of magnitude  in the frustration parameter
$Q/J$,   always in  the weak-frustration  regime  $(Q/J<<1)$.  In what
follows, we take $J=1$.
\begin{figure}[t]
\resizebox{7.5cm}{!}{\includegraphics[angle=270]{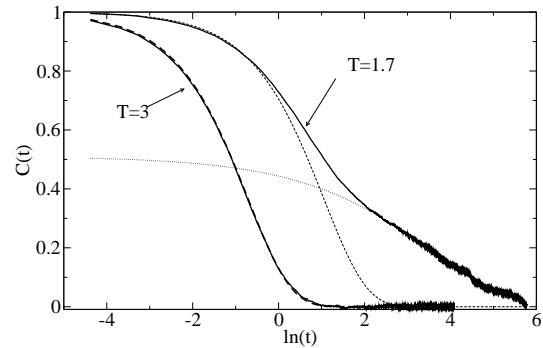}}
\caption{$C(t)$ versus $\ln(t)$ for the
$5-$state clock model  with $Q=0.00625$. Left  curve: $T=3>T^0_c$; the
dashed line,    virtually  indistinguishable  from the  data,    is  an
exponential.  Right  curve: $T=1.7  <T^0_c$;  the  dashed curve  is  an
exponential  and the  dotted curve  is  a stretched exponential  with $\beta
=0.4$.}
\label{fig:2}
\end{figure}

The evolution of  $C(t)$ with $T$  is illustrated in  Fig.~\ref{fig:1}
for  the $5$-state   clock ($Q=0.1$ and   $Q=0.00625$), the $11-$state
clock ($Q=0.05$),  and the  Ising  ($Q=0.001$) models.  At the highest
frustration, illustrated  by Fig.~\ref{fig:1}a,  the  decay of  $C(t)$
appears to proceed in a single step at all  $T$'s. For lower frustrations,
a  $2-$step decay develops as  $T$  is lowered (see Fig.~\ref{fig:1}b
and d).  At  high $T$'s,  typically above the   critical point  of the
coresponding unfrustrated model ($T_c^0\simeq  2.1$  and $T_c^0\simeq 4.51$  for
the $5-$state   clock model and  the  Ising  model, respectively), the
whole time   dependence  of   $C(t)$ is   well   fitted  by a   simple
exponential,  $\exp(-t/ \tau_0(T))$; this  is shown  in Fig.~\ref{fig:2},
but is true for all models and all frustrations.   At low $T$'s, below
$T^0_c$, it is impossible to describe the entire decay  of $C(t)$ by a
single exponential;  as illustrated in Fig.~\ref{fig:2},  the emerging
second  step of  the   relaxation  can be  described   by  a stretched
exponential, $\exp(-(t/ \tau_{KWW}(T))^{\beta(T)})$, while the first step can
still be fitted   by  a simple exponential,   $\exp(-t/ \tau_0(T))$. This
non-exponential  behavior and emergence  of a  two-step decay, both of
which  become  more marked as $T$   is decreased, are  typical of many
actual  glassforming    systems,  especially  the  fragile supercooled
liquids\cite{EAN96}.   One may    however notice that   the  timescale
separation between the  two relaxation steps that  can  be achieved in
the simulations is not sufficient to observe the development of a true
plateau at intermediate times.

\begin{figure}[t]

\resizebox{7.5cm}{!}{\includegraphics{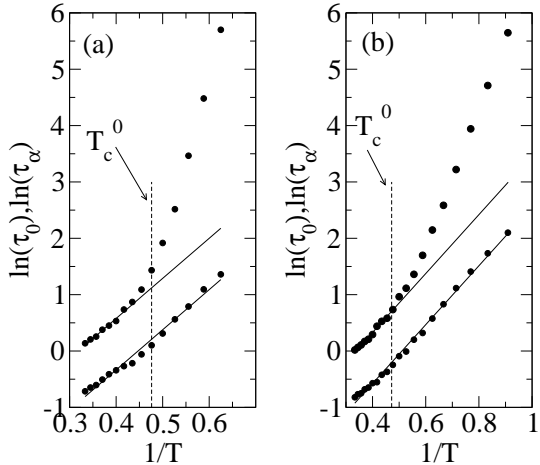}}
\caption{Arrhenius plots of $\tau_0$ (lower curve) and $\tau_\alpha$  (upper
curve)   for the  $5-$state   clock  model  with $Q=0.00625$  (a)  and
$Q=0.05$ (b).}
\label{fig:3}
\end{figure}

To study in more detail the characteristics of the slowing down of the
relaxations  as one  lowers   $T$,  a phenomenon  already   visible in
Fig.~\ref{fig:1},   we have  monitored  as   a   function of $T$   the
relaxation times characteristic of the short- and long-time behaviors,
$\tau_0(T)$ and $\tau_\alpha (T)$ respectively.  The former  is obtained from the
fit to a simple exponential (see above), but an equivalent behavior is
obtained  from the condition $C(\tau_0)=0.9$.  The  latter is defined, as
often  for spin systems,  by the condition  $C(\tau_\alpha)=0.1$; in this case
indeed,  there was too  much uncertainty  in using the three-parameter
stretched exponential   fit that also  requires the  introduction of a
lower  cut-off time.  The  results are illustrated in Fig.~\ref{fig:3}
for the $5$-state clock model. We have plotted  the logarithm of $\tau_0$
and  $\tau_\alpha  $ versus $1/T$.   One  can see that  over  the range of $T$
studied, the  dependence  of $\tau_0(T)$ is  essentially  Arrhenius-like,
i.e., linear   on the diagram, $\tau_0(T)\simeq \tau_{0,\infty}\exp(E_\infty  /T)$, whereas
the dependence of  $\tau_\alpha (T) $ shows  a marked deviation from Arrhenius
behavior below some crossover  temperature in the vicinity of $T^0_c$.
Note   that at high    $T$'s  (above $T^0_c$),  $C(t)$ is  essentially
exponential and the two  times $\tau_0 (T) $  and $\tau_\alpha (T)  $ differ by a
trivial constant, so that in this $T$  range the Arrhenius dependences
of $\tau_0  $  and  $\tau_\alpha  $ are   the same.   At lower  $T$'s,  $\tau_0 $ is
characteristic  of a   ``secondary'' or  ``precursor'' relaxation; one
expects it  to be associated with  a local, weakly or non cooperative,
mechanism of relaxation.  On the  other hand, the super-Arrhenius rise
of $\tau_\alpha $ versus $1/T$ is indicative of cooperative behavior.

\begin{figure}[h]
\resizebox{7.5cm}{!}{\includegraphics{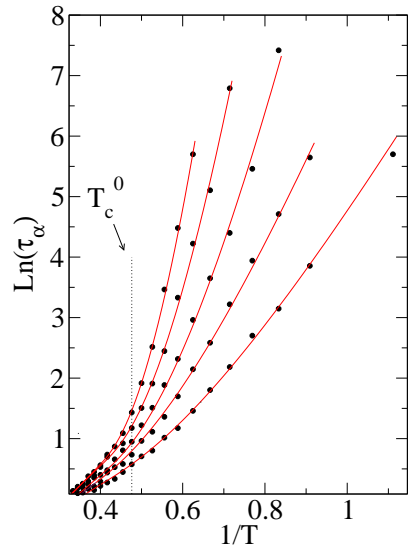}}
\caption{Arrhenius plot of $\tau_\alpha(T)$ for the $5-$state
clock model;   from left  to right:  $Q=0.00625$, $Q=0.0125$, $Q=0.25$,
$Q=0.05$, $Q=0.1$. The full lines are the best fits to Eqs.~(2,3).}
\label{fig:4}
\end{figure}

\begin{figure}[h]

\resizebox{7.5cm}{!}{\includegraphics{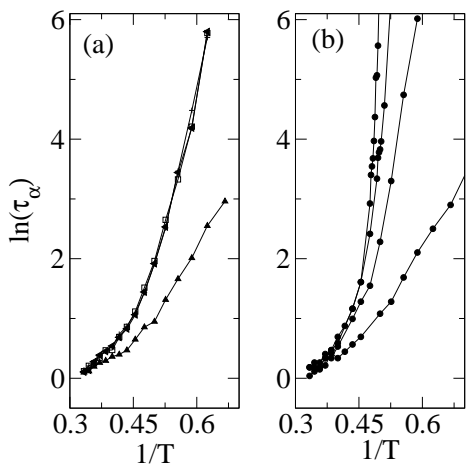}}
\caption{Arrhenius plot of  $\tau_\alpha$   for the  $5-$state   clock  model
and for   lattice   sizes $L=5,10,15,20$  (from  right  to  left): 
$Q=0.00625$ (a) and  $Q=0$ (b).}
\label{fig:5}
\end{figure}

The crossover from  Arrhenius to super-Arrhenius  behavior of $\tau_\alpha$ is
shown  in Fig.~\ref{fig:4} for  the   $5-$state clock  model for   $5$
different frustrations from   $Q=0.1$ down to $Q=0.00625$.  A  similar
trend is  observed for all other  models studied (for the Ising model,
see Ref.\cite{GTV01a,Grous2001})).  Several  points are worth  noting:
first,   such a super-Arrhenius behavior   is  typical of the  viscous
slowing  down   of fragile glassforming   liquids,  the more fragile a
liquid the more  pronounced the super-Arrhenius character\cite{EAN96};
secondly, the  crossover occurs in  the vicinity of the critical point
of the   unfrustrated system; thirdly,   the  departure from Arrhenius
behavior,  i.e., the ``fragility'', becomes  more marked as frustration
decreases: see Fig.~\ref{fig:5}.  To  our  knowledge,  such   a  ``tunable'' fragility  that
provides  a description   over  the  whole spectrum of    glassforming
behavior, from   strong  to  very fragile,    is shown by   no   other
microscopic model.

The competition of interactions acting on very different length scales
can thus   generate fragile  glassforming  behavior   similar to  that
observed  in actual  supercooled liquids.   In the FLDT\cite{KKZNT95},
this  behavior is  attributed to  the  proximity of the critical point
$T^0_c$ of the unfrustrated  system, a critical  point that is avoided
in  the     presence     of    frustration.       Avoided     critical
behavior\cite{KKZNT95,CEKNT96}   results in  the  appearance  at $T$'s
below $T^0_c$ of domains  whose size  is  limited by the  frustration.
These domains are responsible for the non-exponential character of the
relaxation and   the  super-Arrhenius,  activated   $T-$dependence  of
$\tau_\alpha$. Scaling arguments  predict  for  instance  that  the  effective
activation free energy for the $\alpha$ relaxation behaves as
\cite{KKZNT95}

\begin{eqnarray}
E_{\alpha}(T)&=&E_{\infty},   \mbox{\hspace{0.5cm}}    \mbox{\rm     $T>T^*$},\\
&=&E_{\infty}+BT^*\left(1-\frac{T}{T^*}\right)^{8/3},\mbox{\hspace{0.5cm}}   \mbox{\rm
$T<T^*$},
\end{eqnarray}
where $E_{\alpha}(T)=T\ln\left(\frac{\tau_{\alpha}(T)}{\tau_{\infty}}\right)$, with $\tau_{\infty}$
a characteristic  high-T time, and  $T^*$  is close  to  $T^0_c$.  The
existence of  a   crossover in the    vicinity  of $T^0_c$  is  indeed
supported  by the Monte Carlo  results.   More quantitatively, we have
compared the above  prediction with the  simulation data. In doing so,
one must remember that  the range of   $\tau_\alpha$'s spanned by our  data is
limited, as  in most simulations of  glassforming systems, and  is far
less than the   $15$  decades observed experimentally  in  supercooled
liquids.   The high-$T$ parameters $\tau_\infty$  and  $E_\infty$ are independently
obtained  by  an Arrhenius  fit to $\tau_0(T)$   (see  above, where $\tau_\infty/
\tau_{0,\infty}=-\ln(0.1)$ due  to   the     definition   of $\tau_\alpha$
and $\tau_0$), and one is left  with two adjustable parameters, $B$ and
$T^*$, for each frustration (and each model). The resulting curves are
shown  in Fig.~\ref{fig:4}  for   the $5-$state clock   model. For all
models the  agreement   between theory and simulation    is  good.  As
predicted by the theory,   $T^*$   is found  close to   $T^0_c$  (e.g,
$T^*/T^0_c\simeq 1.14-1.20$ for the $5$-state clock and $1.04-1.12$ for the
Ising  model) and the super-Arrhenius  parameter $B$ varies roughly as
$Q^{-1}$.  (As already stressed,  small  frustration $Q$ implies large
super-Arrhenius character and large fragility.)

We  have thus found  that our  simulation data  are compatible with an
activated-type  expression, $\tau_\alpha(T)\propto  \exp(E_\alpha  (T)/T)$ with  $E_\alpha(T)$
given  by the FLDT   expression.  (As discussed in \cite{SW00,GTV01a},
the data are also compatible with $E_\alpha (T)$  described by the entropic
droplet/random-first-order transition approach (EDA)\cite{KTW89,SW00},
which offers a unique  opportunity for comparing these two alternative
theories in more detail.) On the other hand, we have checked that they
are  poorly described   by   a  power-law  dependence,  $\tau_\alpha(T)\propto   |T-
T_c|^{-\gamma}$, except of course  when $Q=0$\cite{Grous2001}.  In both the
FLDT and the EDA, the  slowing down of the relaxation,  as well as the
non exponential character is  attributed to genuinely non-perturbative
effects,   such    as heterogeneities   (domains   or    droplets) and
cooperative, activated processes.     This is   quite different   than
standard  critical slowing   down, as could  occur  if  driven by  the
approach  of a spinodal.  In  particular,  the length scale associated
with the slowing down is predicted to grow slowly with decreasing $T$,
at a much slower rate than the relaxation time  itself. To check this,
we have performed a finite-size study of the $5$-state clock model for
$Q=0.00625$.    As seen   in  Fig.~\ref{fig:5}a,  $\tau_\alpha(T)$   is indeed
unchanged when one  decreases the linear size of  the system from $20$
to  $10$; on the other  hand, it is  strongly altered  when passing to
$L=5$. This seems to confirm that any putative length scale associated
with the strong slowing down is rather modest (of the  order of $5$ or
so)  over the  whole $T-$   range studied.   For comparison, the  same
finite-size  study for the   unfrustrated  ($Q=0$) model, showing  the
usual pattern associated with  critical slowing down, is displayed  in
Fig.~\ref{fig:5}b.

In  summary, due to  their  properties (absence of quenched  disorder,
physical     motivation    in    terms    of  structural  frustration,
$3-$dimensional,  finite-range character,  tunable   fragility),  whose
combination    is found in  no other   models\cite{CLMYB}, the Coulomb
frustrated systems appear as  promising minimal theoretical  models for
describing the slowing down of relaxations in supercooled liquids.

We thank D. Kivelson and L. Berthier for many stimulating discussions.

\end{document}